\documentclass[manuscript]{aastex}

\usepackage{graphicx}

\makeatletter

\begin{document}

\title{Peering into the Giant Planet Forming Region of the \\ TW Hydrae Disk with the Gemini Planet Imager}

\author
{Valerie A. Rapson\altaffilmark{1,2},
 Joel H. Kastner\altaffilmark{1,3}
 Maxwell A. Millar-Blanchaer\altaffilmark{4},
Ruobing Dong\altaffilmark{5}
}

\email{var5998@rit.edu}
\altaffiltext{1}{School of Physics and Astronomy and Laboratory for Multiwavelength Astrophysics, Rochester Institute of Technology, 1 Lomb Memorial Drive, Rochester, NY 14623-5603, USA}
\altaffiltext{2}{Dudley Observatory, 15 Nott Terrace Heights, Schenectady, NY 12308, USA}
\altaffiltext{3}{Chester F. Carlson Center for Imaging Science, Rochester Institute of Technology, 1 Lomb Memorial Drive, Rochester, NY 14623-5603, USA }
\altaffiltext{4}{Department of Astronomy and Astrophysics, University of Toronto, ON, M5S 3H4, Canada }
\altaffiltext{5}{Department of Astronomy, University of California at Berkeley, Berkeley, CA 94720, USA}

\begin{abstract}
We present Gemini Planet Imager (GPI) adaptive optics near-infrared images of the giant planet-forming regions of the protoplanetary disk orbiting the nearby ($D=54$ pc), pre-main sequence (classical T Tauri) star TW Hydrae. The GPI images, which were obtained in coronagraphic/polarimetric mode, exploit starlight scattered off small dust grains to elucidate the surface density structure of the TW Hya disk from $\sim$80 AU to within $\sim$10 AU of the star at $\sim$1.5 AU resolution. The GPI polarized intensity images unambiguously confirm the presence of a gap in the radial surface brightness distribution of the inner disk. The gap is centered near $\sim$23 AU, with a width of $\sim$5 AU and a depth of $\sim$50\%. 
In the context of recent simulations of giant planet formation in gaseous, dusty disks orbiting pre-main sequence stars, these results indicate that at  least one young planet with a mass $\sim$0.2 $M_J$ could be present in the TW Hya disk at an orbital semi-major axis similar to that of Uranus. If this (proto)planet is actively accreting gas from the disk, it may be readily detectable by GPI or a similarly sensitive, high-resolution infrared imaging system.

\end{abstract}


\section{Introduction}
Recent numerical simulations of planet formation processes in disks orbiting young stars demonstrate that such planet-building activity should leave distinct imprints on disk dust density structures \citep[e.g.,][]{Zhu2014,Dong_etal2015b,Dong_etal2015a}. Planet-disk interactions cause material that is co-orbiting with a massive planet to be transported outward via deposition of angular momentum onto the disk, creating pressure gradients that are manifested in disk density structures such as gaps, rings or spirals \citep[e.g.,][]{Bryden1999,Pinilla2012,Dong_etal2015b,Dong_etal2015a}.  Many such examples of ring/gap features have now been detected via submm-wave interferometry within disks orbiting young (age $<$10 Myr), low-mass stars \citep[e.g.,][]{Hughes2007,Andrews2009,Isella2010,Andrews2011,Rosenfeld2013}. 
With the advent of extreme adaptive optics (EAO) near-infrared polarimetric imaging instruments, such as the Gemini Planet Imager (GPI) on Gemini South and SPHERE on the Very Large Telescope, we can probe as close or closer to the central stars --- within a few AU --- by imaging light scattered off micron-sized (or smaller) dust grains \citep[e.g.,][]{Garufi2013, Rapson2015,Thalmann2015}. Comparison of such near-infrared, polarimetric images with similarly high-resolution ALMA submm imaging of thermal emission from dust \citep[e.g.,][]{ALMA2015} potentially offers a powerful means to characterize the radial and azimuthal dust density distributions within giant planet formation regions of protoplanetary disks and, perhaps, to pinpoint the locations of young planets \citep{Zhu2014,Dong_etal2015a}.

Thanks to its combination of proximity \citep[D$=$54 pc;][]{Torres2008}, advanced age \citep[$\sim$8 Myr;][]{Ducourant2014}, and nearly face-on viewing geometry \citep[$i\approx7^\circ$;][]{qi2004}, the disk orbiting TW Hya represents particularly fertile ground for such searches for evidence of planet formation.  TW Hya is a $\sim$0.8 M$_{\odot}$ star of optical spectral type K6 \citep{torres2006}. Its relatively gas-rich, massive \citep[$\sim$0.05 $M_\odot$;][]{Bergin2013} circumstellar disk has been imaged over a broad range of wavelengths, revealing rich disk structure and chemistry. \citet{andrews2012} and \cite{Rosenfeld2012} imaged TW Hya with the Submillimeter Array and ALMA and found its CO gas disk extends to $\gtrsim$215 AU, similar to the radial extent of scattering of starlight by small (micron- and submicron-sized) dust grains \citep{Weinberger2002,Apai2004}, while the larger (mm-sized) dust grain population traced by 870 $\mu$m continuum emission displays a sharp outer edge at 60 AU. \citet{debes2013} used HST/STIS coronagraphic imaging to trace starlight scattered off dust grains in the outer disk and find a gap in the disk at $\sim$80 AU that they contend could be opened by a planetary companion of a few hundredths of a Jupiter mass. A ring of C$_2$H emission extending from $\sim$45 AU to $\sim$120 AU overlaps this 80 AU gap and may trace a region of highly efficient photodissociation of hydrocarbons in the surface layers of the disk \citep{kastner2015}. Emission from N$_2$H$^+$ imaged with the SMA reveals an inner hole at R$\lesssim$30 AU that likely traces the inner edge of the CO ``snow line,'' where CO molecules freeze out onto dust grains \citep{qi2013}. \citet{akiyama2015} identified a possible gap in the disk interior to this snow line, at R$\sim$20 AU, via polarimetric near-infrared imaging of light scattered off dust grains with Subaru/HiCIAO. Modeling of the spectral energy distribution of the disk \citep{calvet2002,menu2014} further indicates the presence of a cavity interior to R$\lesssim$4 AU. Within this innermost disk region, the disk evidently harbors a few tenths of a lunar mass in submicron-sized dust grains \citep{Eisner2006}.

Here, we present EAO near-IR coronagraphic/polarimetric images of the TW Hya disk that we obtained with the Gemini Planet Imager \citep[GPI;][]{Macintosh2008,Macintosh2014} on the Gemini South telescope. These images directly probe the disk dust structure, via the intensity of scattered incident starlight, to within $\sim$10 AU of TW Hya with unprecedented linear spatial resolution of $\sim$1.5 AU. The GPI imaging observations hence allow us to study the disk structure within the giant planet forming region around TW Hya, complementing and elaborating on the aforementioned recent mm-wave interferometric, polarized near-infrared, and optical imaging data obtained for this heavily scrutinized protoplanetary disk. 

\section{Observations and Data Reduction} 

Gemini/GPI polarimetric images of TW Hya were obtained through $J$ (1.24 $\mu$m) and $K1$ (2.05 $\mu$m) band filters and 0.184$''$ and 0.306$''$ diameter coronagraphic spots on April 22, 2015. Eight (four-image) sets of $J$ band images and 11 sets of $K1$ band images were obtained at waveplate position angles of 0$^\circ$, 22$^\circ$, 45$^\circ$, and 68$^\circ$ with exposure times of 60 s each. The $K1$ band images were obtained first, and over the course of the observations, the airmass and DIMM seeing ranged from 1.018$"$ to 1.034$"$ and 1.66$"$ to 0.8$"$ , respectively. Under these conditions, GPI's EAO system delivers nearly diffraction-limited imaging, with angular resolution of $\sim$0.032$''$ and $\sim$0.055$''$ at $J$ and $K1$, respectively, on the 8 m of the Gemini South telescope (which has an effective pupil aperture of  7.77 m). Due to the target's relatively faint I magnitude \citep[$I \sim 9.2$;][]{barrado2006}, the EAO system had difficulties keeping the focal plane mask properly aligned; only 25 of the (44) $K1$ image frames were included in the final reduction. 

Polarimetric images in each filter were reduced using the GPI pipeline v1.3.0 \citep{Maire2010, Perrin2014_pipe}, following methods similar to those described in \citet{Perrin2014} and \citet{Rapson2015}. To summarize, the images were cleaned by removal of correlated noise due to detector readout electronics and microphonics. We then performed interpolation over bad pixels, and subtracted background images from the target $K1$ frames. Calibration spot grids, which define the location of each polarization spot pair produced by the lenslet array, were used to extract intensity data from each raw image and produce a pair of orthogonally polarized images. Satellite spots on both the $J$ and $K1$ band images were used to determine the location of the the central star behind the coronagraph. The cubes were cleaned using a double differencing routine, and then instrumental polarization was subtracted as described in \citet{MillarBlanchaer2015}. Each image was then smoothed using a Gaussian filter with a FWHM of $\sim$2.3 pixels ($\sigma$ = 1 pixel) to reduce pixel to pixel noise. The cubes were all registered to place the obscured star at the center, and were then rotated to the same sky coordinate alignment (i.e., N up and E to the left). The orthogonally polarized images obtained at the four different waveplate angles were combined via the GPI pipeline to produce a Stokes data cube with slices $I$, $Q$, $U$ and $V$. Lastly, the radial and tangential Stokes parameter images $Q_r$ and $U_r$ were calculated via the pipeline from the Stokes cube \citep[as in][albeit with the opposite sign convention]{schmid2006}. Assuming that all the polarized flux is in the tangential component (which is expected for stellar photons single-scattered off circumstellar material), calculation of $Q_r$ avoids the positive bias induced by calculating total polarized intensity as the sum of the squares of the Stokes $Q$ and $U$ images. As absolute flux calibration of GPI polarized intensity images remained uncertain as of the writing of this paper, the $Q_r$ images are presented here in instrumental units.

 \section{Results}
 
 In Fig.~1 we display total polarized intensity ($Q_r$) and radially scaled polarized intensity ($Q_r \times R^2$) images of the TW Hya disk. The total intensity images ($I$) are dominated by the residual PSF, and the disk was not detected; indeed, due to its azimuthal symmetry, the disk is indistinguishable from the stellar PSF in  unpolarized angular differential imaging. The $Q_r$ images trace the intensity of scattered starlight as a function of position, while the radially scaled $Q_r$ images account for the dilution of incident starlight and thereby better represent the dust density distribution across the disk surface \citep[e.g.,][]{Garufi2014,Rapson2015}. The total polarized intensity images reveal a bright central ring with relatively sharp outer edge near $\sim$0.3$"$ ($R\sim18$ AU), a deficit of scattered light just outside this radius, and a fainter scattering halo extending to at least $\sim$1$"$ ($R\sim55$ AU). All of these features are seen at both $J$ and $K1$, and are especially evident in the scaled polarized intensity images  (righthand panels of Fig.~1). The latter images also appear to show an inner dust cavity within $\sim$15 AU, although this region of the images lies very near the inner working angle of the coronagraph (see below). The scattered light from the disk is detected at higher signal-to-noise ratio at $J$ band  because of the combination of lower background and brighter incident radiation field, and, possibly, greater dust scattering efficiency at this wavelength. 

In Fig.~2, we present radial brightness profiles obtained from the $J$ and $K1$ band polarized intensity and scaled polarized intensity images along directions parallel and perpendicular to disk position angle, i.e., the line marking the intersection of the disk equatorial plane and the plane of the sky, as inferred from CO kinematics \citep[151$^{\circ}$, measured east from north;][]{Rosenfeld2012}. The radial profiles obtained from the unscaled $J$ and $K1$ polarized intensity images (lefthand panels of Fig.~2) display a sharp dip in surface brightness near $\sim$20--25 AU, as well as a weaker inflection near $\sim$30 AU. The profiles furthermore indicate that the outer scattered light halo drops smoothly out to $\sim$100 AU, a displacement roughly corresponding to the corner of the field of view of GPI at  $J$ band. The radial profiles of the scaled polarized intensity at $J$ and $K1$ (righthand panels of Fig.~2) more clearly show the sharp dip feature (hereafter referred to as a gap) to be centered at $R\sim23$ AU; its FWHM of $\sim$5 AU indicates that the gap is well resolved by GPI. 

The radial profiles in Fig.\ 2 also indicate that the polarized intensity is relatively isotropic, with large-scale asymmetries only of order $\sim$10\% and no clear systematic differences in surface brightness along directions parallel vs.\ perpendicular to the disk position angle. This is consistent with the modest disk inclination determined from radio line data \citep[$i \approx 7^\circ$;][and refs.\ therein]{Rosenfeld2012}. In the $K1$ image, the inner disk surface brightness appears somewhat brighter to the southeast. Given that the same asymmetry is not present in the $J$ image, however  (compare left panels of Figs.\ 1 and 2), the apparent inner-disk asymmetry at $K1$ is most likely caused by poor centering of the star+disk under the coronagraph.

In Figure 3, we present surface brightness profiles obtained by averaging the $J$ and $K1$ polarized intensity images over concentric elliptical annuli  with minor:major axis ratios of 0.99 (approximating a circular disk with inclination of 7$^\circ$) at single-pixel (0.014$"$) intervals, with the ellipses oriented at a position angle of $151^\circ$. These profiles, which are discussed in detail in \S 4, decline rapidly from $R\sim12$ AU  to the gap feature at $R\sim20$--25 AU, appear nearly flat out to $\sim$40 AU, and then drop precipitously beyond this radius. There is also an inflection in surface brightness at $R\sim80$--90 AU that is more apparent in the (higher S/N ratio) $J$ band radial profile. 

The profiles in Figs.\ 2, 3 furthermore suggest that the bright inner disk has a ring-like structure; i.e., the surface brightness of polarized intensity peaks at $\sim10$--15 AU, with a potential decline interior to this radius. We note, however, that the peak in polarized intensity appears to lie closer to the star at $J$ than at $K1$ (peaks near $\sim$10 AU and $\sim$12 AU, respectively). This suggests that the decrease in polarized intensity near the inner working angle of the coronagraph may be instrumental in origin, although there remains the possibility that there is in fact a deficit of small dust grains in the TW Hya disk within $\sim$10 AU of the star. If real, this inner cavity dimension would be somewhat larger than previously deduced from SED fitting \citep[which yielded an inner cavity size scale of $\sim$3--4 AU;][]{calvet2002,menu2014}. 

\section{Discussion}

\subsection{The structure of the TW Hya dust disk from $\sim$10 AU to $\sim$80 AU}

Previous near-infrared coronagraphic imaging of TW Hya disk was conducted with HiCIAO, in polarimetric mode, on the 8 m Subaru telescope \citep[][]{akiyama2015}. This imaging revealed a potential gap at a radial position of $R\sim20$ AU in the scattered-light surface brightness distribution of the TW Hya disk. This feature is readily apparent in the radial polarized intensity profiles extracted from our GPI coronagraphic/polarimetric imaging (Figs.\ 2, 3). The GPI radial profiles furthermore provide refined measurements of the position and width of the inner gap detected by Subaru/HiCIAO, as well as the dependence of scattered light on radius from $\sim$10 AU to $\sim$80 AU. Specifically, we find the scaled polarized intensity (Fig.\ 2, right panels) displays a clear local minimum at $R\sim23$ AU with a FWHM of $\Delta R\sim5$ AU and depth of $\sim$50\%. In the azimuthally averaged radial profiles (Fig.\ 3), this feature has a bowl-shaped appearance. The GPI data also appear to confirm the presence of a gap near $R\sim80$--90 AU that was inferred from HST imaging \citep{debes2013}, in the form of a weak inflection in the GPI $J$ band surface brightness profiles. We caution, however, that this radius roughly corresponds to the limit of the GPI field of view at $J$.

In characterizing the potential gap at $\sim$20 AU, \citet{akiyama2015} parameterized the azimuthally averaged polarized intensity of the TW Hya disk in terms of a ``stair-like'' decline in radial power law ($r^\gamma$), with a drop from $\sim$10 to $\sim$20 AU characterized by $\gamma \approx -1.4$, a flattening to $\gamma \approx -0.3$ between $\sim$20 AU and $\sim$40 AU, and then a turnover to $\gamma \approx -2.7$ out to $\sim$80 AU. Adopting the same ``three-zone'' model (Fig.\ 3, top panel), we find similar values of $\gamma \approx -1.7$ and $\gamma \approx -0.4$, respectively, in the two zones within $\sim$40 AU, but we measure a considerably steeper decline of $\gamma \approx -3.9$ beyond $\sim$40 AU. 

However, the clear appearance of the gap-like feature at $\sim$23 AU in Fig.\ 2 suggests it is preferable to parameterize the azimuthally averaged polarized intensity profiles (Fig.\ 3) in terms of  a smooth decline in the inner disk (from $\sim$10 AU to $\sim$45 AU) that is interrupted by the gap feature at $\sim$23 AU, and a turnover to a steeper decline beyond $\sim$45 AU (i.e., a ``two-zone'' model). Adopting this parameterization, we find the radial dependence of polarized intensity is characterized by $\gamma \approx -1.2$ between $\sim$10 AU and $\sim$45 AU and by $\gamma \approx -3.9$ between $\sim$45 AU and $\sim$80 AU (Fig.\ 3, top). The latter value is rather extreme compared with the value $\gamma \approx -2$ that is predicted for radial scattered-light profiles by simulations that adopt ``conventional'' surface density and scale height profiles \citep[e.g.,][ see next section]{Dong2015}.

\subsection{TW Hya and V4046 Sgr: similarities and differences}

In the right panel of Fig.\ 3, we compare the radial near-IR polarized intensity profiles of TW Hya with those obtained from GPI imaging of V4046 Sgr \citep{Rapson2015}, a close binary, classical T Tauri system that, like TW Hya, is relatively nearby and evolved ($D \approx 73$ pc; age $\sim$20 Myr). The TW Hya disk and the V4046 Sgr (circumbinary) disk are also similarly massive and chemically rich \citep[][and references therein]{Kastner2014}. The comparison demonstrates that the two disks display similar  slopes of azimuthally averaged polarized intensity in the $\sim$12--30 AU range; both show gaps, in the form of depressions in polarized intensity, near $\sim$20 AU; and both show turnovers to much steeper radial polarized intensity profiles in their outer disks.  For TW Hya, the turnover to a steeper slope occurs at $\sim$45 AU, as noted above, whereas for V4046 Sgr the turnover lies near $\sim$28 AU and the slope of the outer disk profile is steeper \citep[$\gamma \approx -5.5$;][]{Rapson2015}. The gap feature near $\sim$20 AU is evidently also much wider and deeper in the TW Hya disk than in the V4046 Sgr disk.

In our tests of Monte Carlo scattering models, we found that the radial polarized intensity gradient is relatively insensitive to the radial gradients in disk surface density and scale height, within reasonable ranges for these parameters (see \S 4.3). There would hence appear to be two potential explanations for the turnovers in the radial polarized intensity profiles of both disks: shadowing of the outer disk by the inner disk, or a sharp change in the radial gradient of the specific surface density or scale height of small and/or highly-IR-reflective dust grains. With regard to these alternatives, it is intriguing that the radial positions of the turnovers of the polarized intensity profiles in the TW Hya and V4046 Sgr disks at $\sim$45 AU and $\sim$28 AU, respectively, appear to closely coincide with the inner edge of a ring of C$_2$H emission \citep[TW Hya;][]{kastner2015} and the inner edge of a ring of submm-wave continuum emission  \citep[V4046 Sgr;][]{Rapson2015}.  In the case of TW Hya, the ring of C$_2$H may mark a region of enhanced stellar-irradiation-driven desorption of volatiles from grain surfaces, and/or photodestruction of dust grains themselves, in disk surface layers \citep[][]{kastner2015}. In the case of V4046 Sgr, the ring of submm emission is likely due to a zone of dust grain growth and grain size segregation processes that are potentially related to planet building \citep{Rosenfeld2013}. The apparent spatial correlations between the turnovers in radial polarized intensity profiles (Fig.\ 3) and these zones of C$_2$H and large grain production may or may not be significant, but are deserving of further study, alongside further explorations of the radial intensity profiles generated by Monte Carlo disk scattering models (\S 4.3).

\subsection{TW Hya disk gap: clearing by a giant planet?} 

Both the depth ($\sim$0.5 dex) and width ($\sim$5 AU) of the apparent gap at $R\sim23$ AU in the TW Hya disk, as imaged by GPI, appear generally consistent with the predictions of simulations of disk gap clearing by giant planets \citep{Dong_etal2015a}. 
In Figs.\ 4 and 5, we directly compare the GPI polarized intensity images and radial profiles, respectively, with model images and profiles of a protoplanetary disk with an embedded planet of mass 0.16 $M_J$ ($2\times10^{-4}$ $M_\star$) at an orbital semimajor axis of 21 AU. The star/disk/planet model and resulting simulated near-IR images have been generated via the same methodology and Monte Carlo radiative transfer code as is described in detail in \citet{Dong_etal2015a}. To investigate the effect of signal-to-noise (S/N) ratio on the comparisons of model and data, we generated low-noise and high-noise model images consisting of runs with $10^9$ and $10^8$ photons, respectively. The latter runs yield images more closely resembling the S/N ratios of the GPI images near the gap region of the disk. The images were then convolved with an approximate instrumental PSF.  

The similarity of the ring-gap structure in the model and data is evident from Fig.\ 4. In the low-noise model images, a spiral feature resulting from planet-induced resonances in the disk is also apparent both within and outside the darker gap near 20 AU. In the high-noise model images, only the in-gap portion of the spiral is detectable, as the surface brightness enhancement of the outer spiral is modest ($\sim$5\%) compared with that of the in-gap spiral ($\sim$30\%). Such a spiral structure is not apparent in the GPI images, although given the S/N ratio of the GPI images within and around the gap (S/N $\sim$ 10), the outer spiral would not be detectable (middle panels of Fig.\ 4). Such a direct comparison is further complicated by a modest asymmetry in surface brightness at radii $\stackrel{<}{\sim}$25 AU in the GPI images that is due to imperfect centering of the star behind the coronagraph (\S 2).

In Fig.\ 5 we compare the observed and model radial profiles, where the latter have been renormalized in intensity so as to match the GPI data. We find the model intensity renormalizations necessary to match the data in the $J$ and $K1$ bands are the same to within $\sim$20\%, indicating that the \citet{Dong_etal2015a} model well reproduces the wavelength dependence of scattered light. 
It is evident from Fig.~5 that the width, depth, and overall shape of the gap feature in the model disk with embedded 0.16 $M_J$ planet provide a good match match to the GPI data, over the range $\sim$15 AU to $\sim$25 AU.

It is also evident that the model slope is steeper than the data for radii $\stackrel{<}{\sim}$15 AU and $\stackrel{>}{\sim}$25 AU. Under the assumptions that the radial profile of the dust surface density ($\Sigma_{\rm dust}$) and the scale height of the dust in the vertical direction ($h_{\rm dust}$) both obey simple power laws and that the dust population is uniform (i.e., the dust scattering properties do not vary) throughout the disk, we have experimented with disk models having various radial dependences of $\Sigma_{\rm dust}$ and $h_{\rm dust}$. We found that changing $\Sigma_{\rm dust}$ and $h_{\rm dust}/r$ from their original \citep{Dong_etal2015a} radial dependencies of $\Sigma_{\rm dust} \propto 1/r$ and $h_{\rm dust}/r\propto r^{0.25}$ to $\Sigma_{\rm dust} \propto r$ or  $h_{\rm dust}/r\propto r^{0.5}$, respectively, results in a change of a factor of a few in absolute disk brightness in the near-IR. However, these power law parameter changes have relatively little impact on the radial dependence of scattered light; specifically, the radial scattered light power law index $\gamma$ changes by at most $\sim$0.3. This is because scattered light emerges from the disk surface with an intensity that is largely determined by the (grazing) incidence angle of starlight on the surface \citep{Takami2014}; changes to the simple radial power-law dependences of $\Sigma_{\rm dust}$ and $h_{\rm dust}$, while changing the disk flaring angle somewhat, do not yield significant changes to the radial dependence of this grazing incidence angle. 

The foregoing suggests that future work aimed at exploring the discrepancy between the observed and model radial profiles should focus on invoking more complex disk radial and vertical dust density dependences (e.g., broken power laws), potential radial and vertical gradients in the grain scattering properties, and/or disk shadowing effects. The first two possibilities are well motivated by observations at other wavelengths (\S 4.2) as well as by the likelihood that grain growth within disks can produce complicated radial dependencies of $\Sigma_{\rm dust}$ for specific dust grain sizes, superimposed on simple power-law $\Sigma_{\rm dust}$ forms for the disk as a whole \citep{Birnstiel2010,Birnstiel2012,Pinilla2012}. Meanwhile, present models suggest that an outer disk that is completely shadowed by the inner disk can display a very steep scattered light radial power law index \citep[see, e.g., Fig.\ 6 of][]{Dong2015}.

\section{Conclusions}

We have presented GPI polarized intensity images of TW Hya that reveal the surface dust density structure of the disk from $\sim$80 AU to within $\sim$10 AU of the star at $\sim$1.5 AU resolution. The GPI imaging unambiguously confirms the presence of a gap in the disk at $\sim$23 AU, with a width of $\sim$5 AU and a depth of $\sim$50\%. 
The comparison between these GPI data and a Monte Carlo radiative transfer model of a disk with embedded planet (\S 4.3) appears to provide compelling evidence that a (proto)planet is likely responsible for clearing this gap in the TW Hya disk. We caution, however, that other mechanisms may be responsible for generating gap and ring structure in protoplanetary disks. For example, the gap in the TW Hya disk that we have resolved with GPI could be a natural consequence of the radial dependence of grain fragmentation rates \citep{Birnstiel2015}. This hypothesis could be tested via ALMA submm continuum imaging of the TW Hya disk at spatial resolution comparable to that of the GPI imaging. Alternatively, \citet{Zhang2015} have proposed that disk gaps may form at radii corresponding to ice condensation fronts, which act as catalysts for rapid, localized grain growth. Although such a model may pertain to the TW Hya disk, we note that the gap at $\sim23$ AU seen in GPI and Subaru/HiCIAO imaging lies well outside the ($\sim$5 AU) radius where water ice is expected to form on dust grains, and well inside the ($\sim$30 AU) inner edge of a ring of N$_2$H$^+$ submm line emission that is hypothesized to mark the radius where the disk midplane drops to the ($\sim$20 K) temperature at which CO freezes out \citep{qi2013}. 

Followup coronagraphic imaging with GPI and with SPHERE (on the VLT) is hence now warranted, to test the hypothesis that at  least one gas giant planet with a mass of $\sim$0.2 $M_{Jup}$ is actively forming in the TW Hya disk at an orbital semi-major axis similar to that of Uranus. Deeper polarimetric imaging with these instruments could reveal the spiral signatures that result from a massive planet exciting resonances within the disk (Fig.\ 4). Thermal emission from the planet itself would likely test the limits of GPI's detection capabilities in its angular and spectral differential imaging mode \citep[e.g.,][]{Macintosh2015}. However, if the planet is actively accreting disk gas, its accretion luminosity may outshine the planet itself by several orders of magnitude in the near-infrared \citep{Zhu2015}. Such a disk-embedded, accreting protoplanet should be readily detectable in orbit about TW Hya by the present generation of EAO near-infrared imaging systems \cite[e.g.,][]{Sallum2015}.

\acknowledgements
{This work is based on observations obtained at the Gemini Observatory, which is operated by the Association of Universities for Research in Astronomy, Inc., under a cooperative agreement with the NSF on behalf of the Gemini partnership: the National Science Foundation (United States), the National Research Council (Canada), CONICYT (Chile), the Australian Research Council (Australia), Minist\'{e}rio da Ci\^{e}ncia, Tecnologia e Inova\c{c}\~{a}o (Brazil) and Ministerio de Ciencia, Tecnolog\'{i}a e Innovaci\'{o}n Productiva (Argentina). Support is provided by the National Science Foundation grant AST-1108950 to RIT. We thank the anonymous referee for helpful comments.}

\begin{figure}[htbp]
\label{fig:polintensityImages}
\begin{center}
\includegraphics[width=3in]{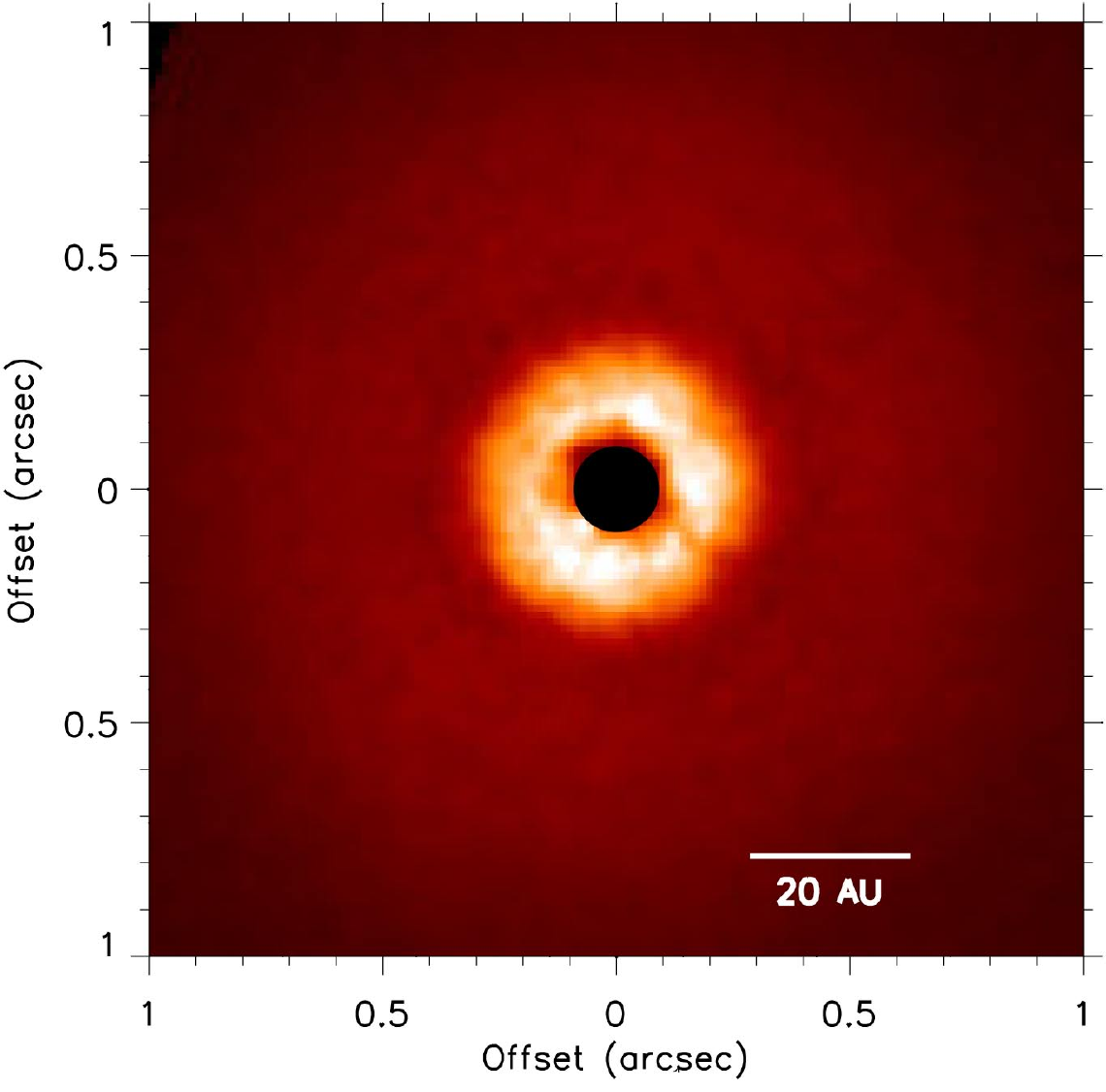} 
\includegraphics[width=3in]{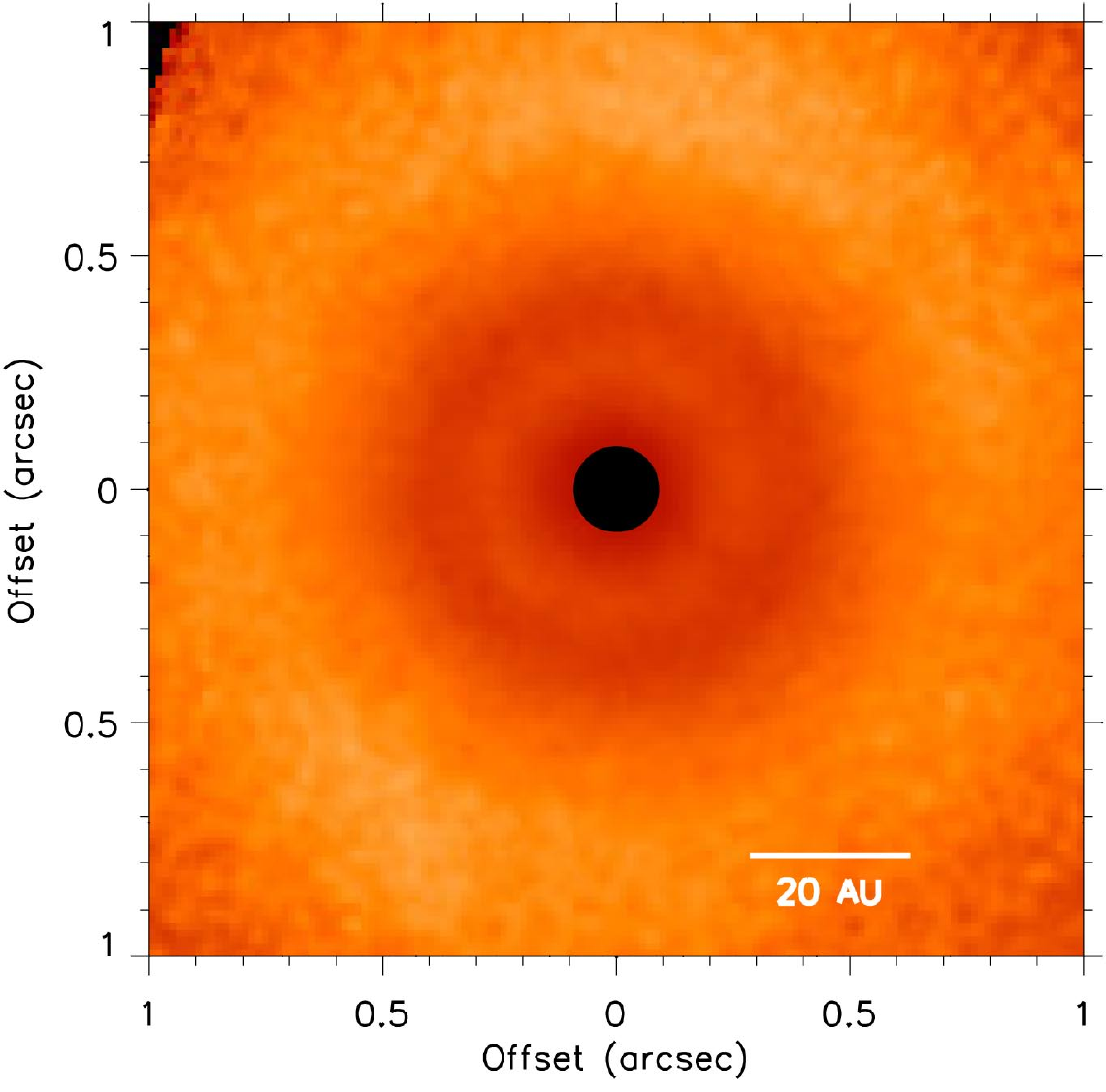} 
\includegraphics[width=3in]{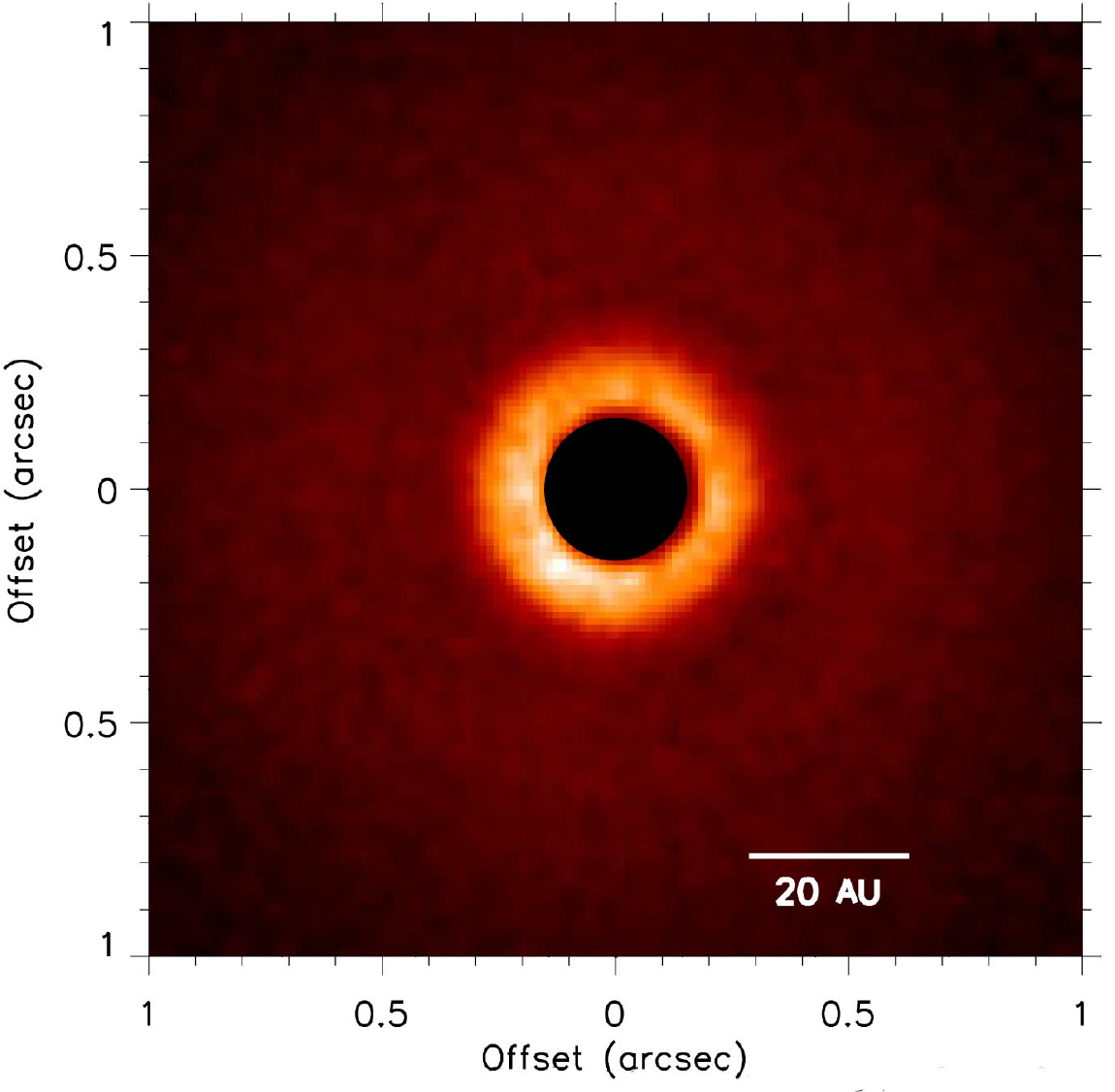} 
\includegraphics[width=3in]{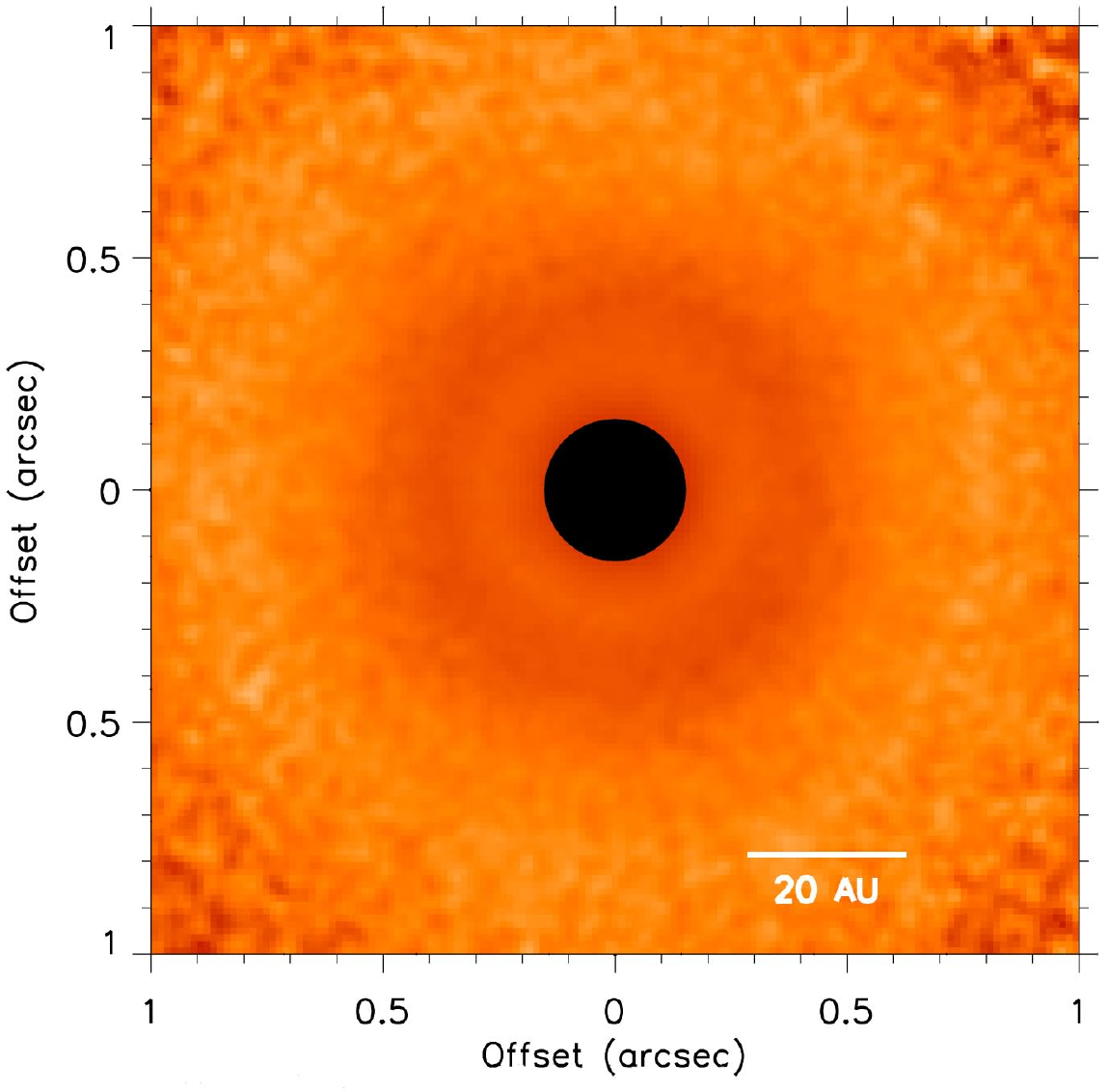} 
\caption{Left: GPI $J$ band (top) and $K1$ band (bottom) polarized intensity ($Q_r$) images of the TW Hya disk. Right: $Q_r(i,j)$ scaled by $r^{2}(i,j)$, where $r(i,j)$ is the distance (in pixels) of pixel position $(i,j)$ from the central star, corrected for projection effects. All images are shown on a linear scale. The coronagraph is represented by the black filled circles and images are oriented with north up and east to the left.} 
\end{center}
\end{figure}

\begin{figure}
\begin{center}
\includegraphics[width=3in]{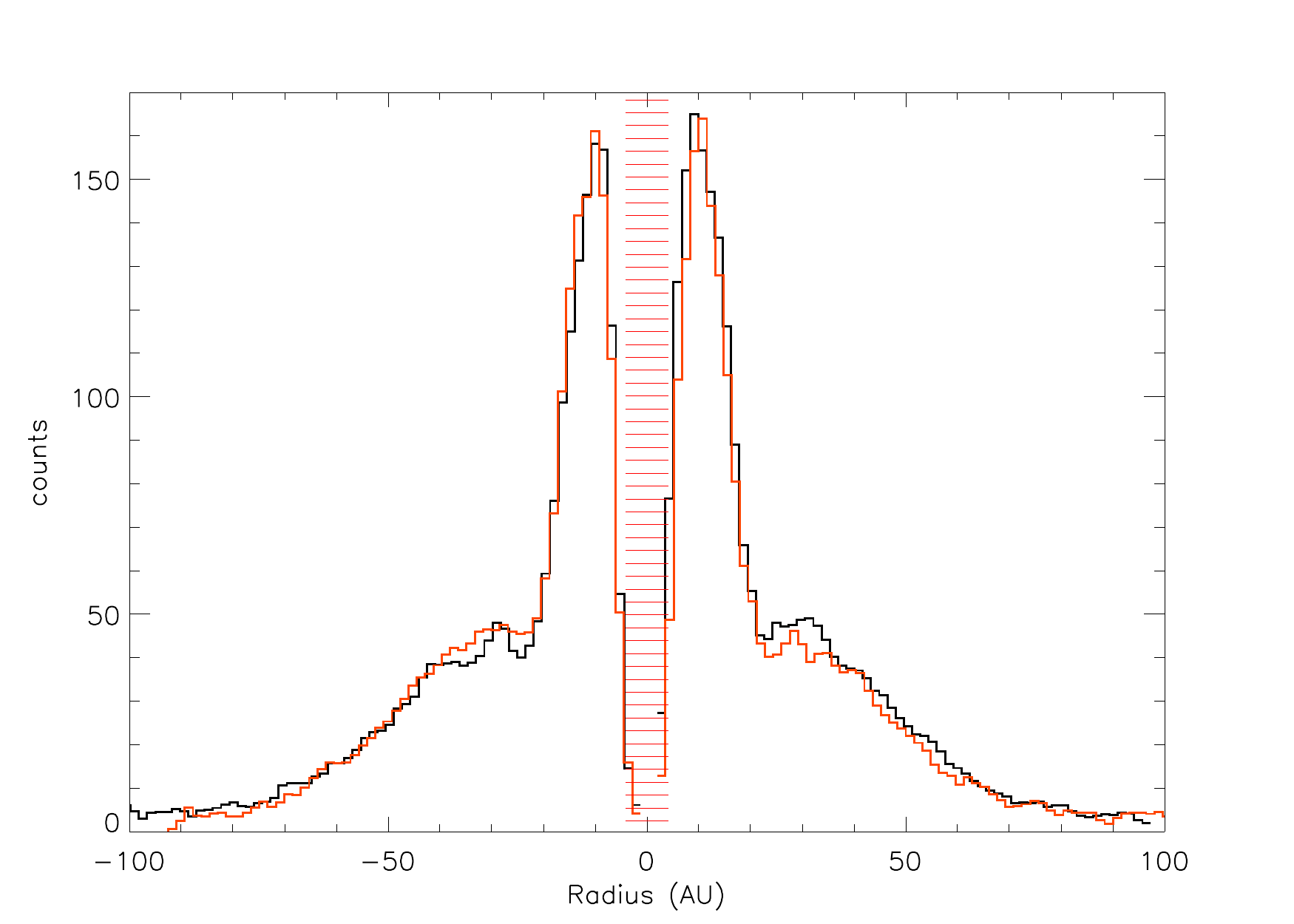}
\includegraphics[width=3in]{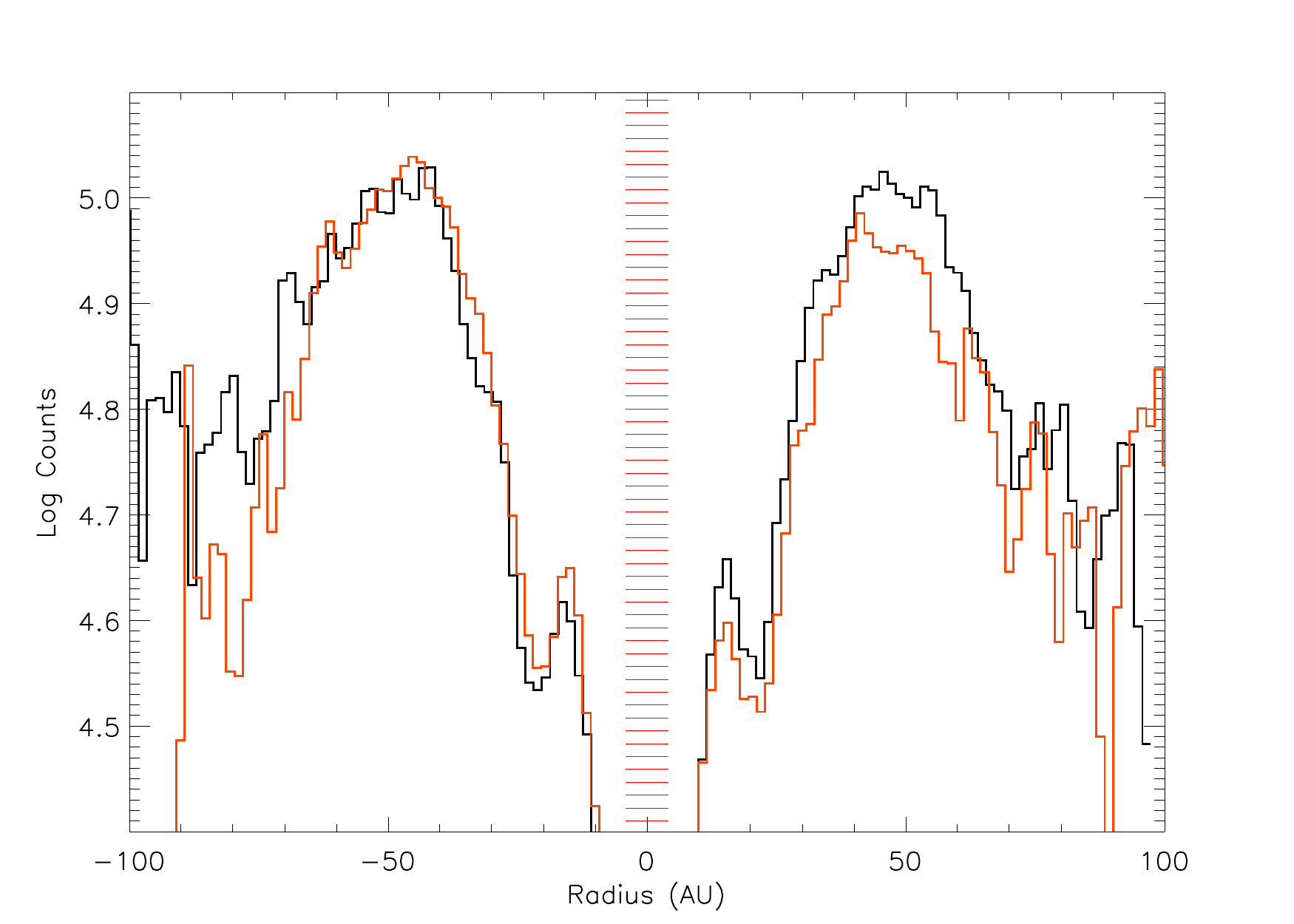}
\includegraphics[width=3in]{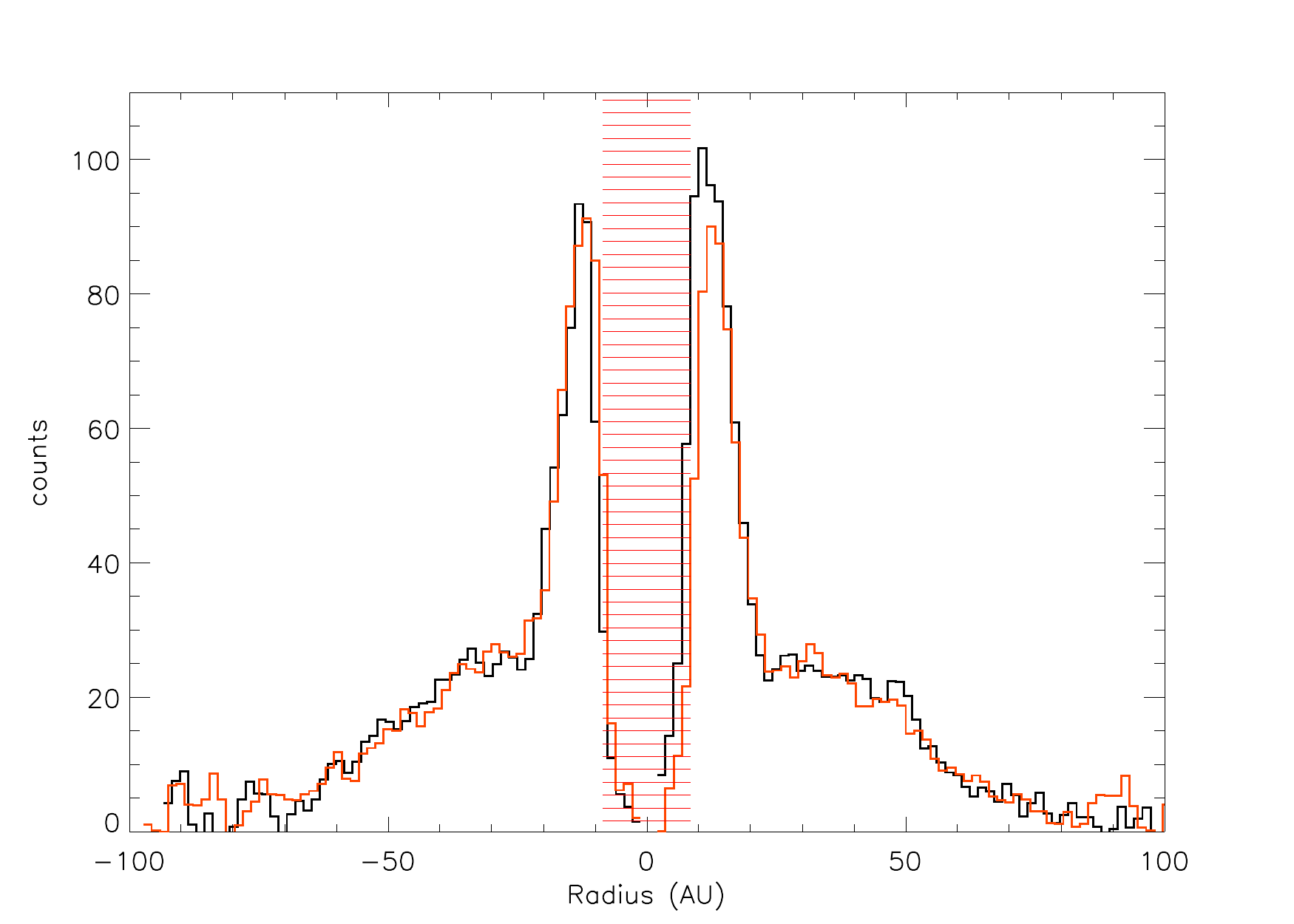}
\includegraphics[width=3in]{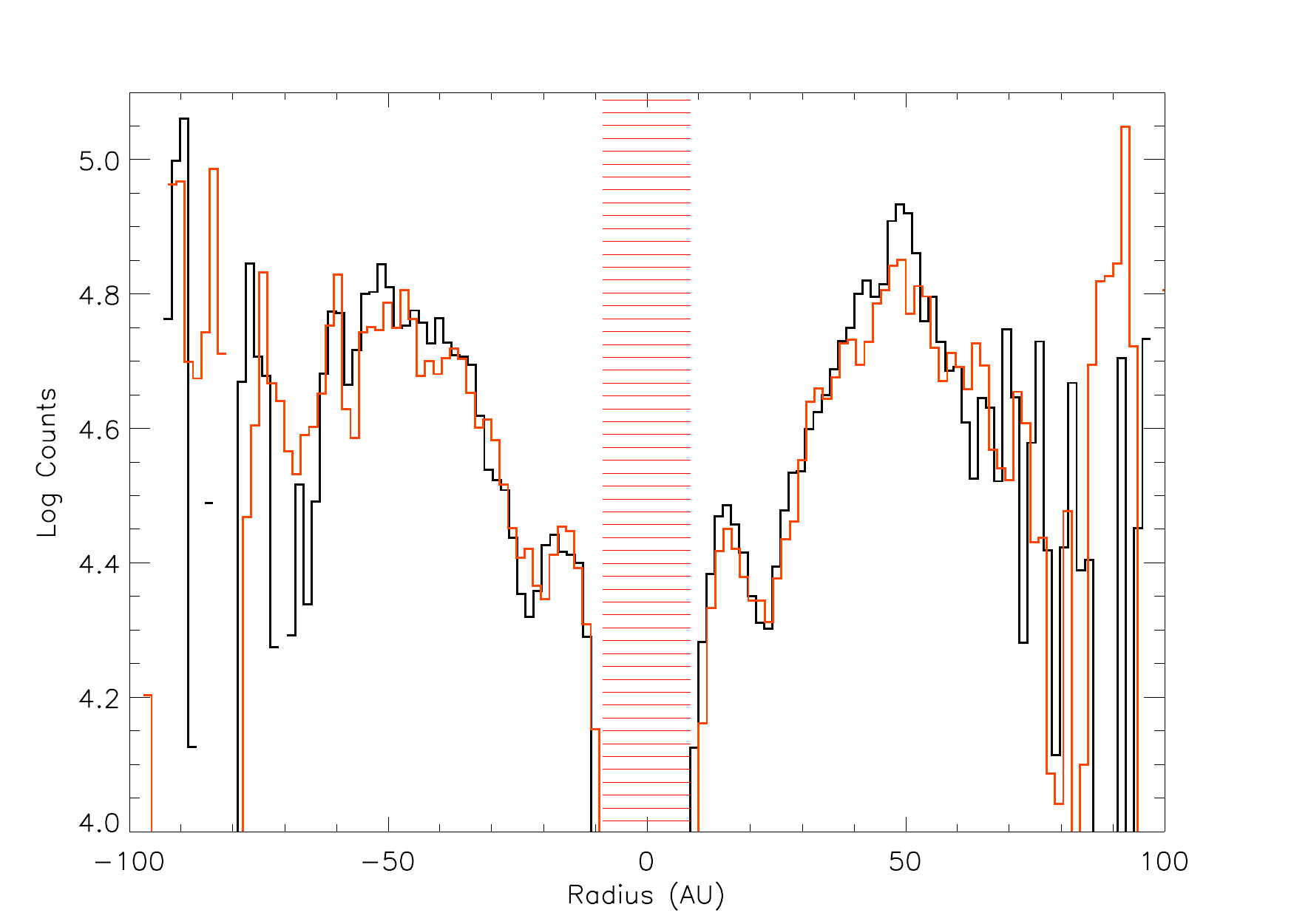}
\caption{Left panels: Background-subtracted radial profiles extracted from the $J$ (Top) and $K1$ (bottom) $Q_{r}$ image, binned 2$\times$2 pixels, parallel (black) and perpendicular (orange) to the 151$^{\circ}$ position angle of the disk's equatorial plane. Right panels: same as the left panels, for the scaled $Q_{r}$ images, without subtracting background. The horizontal red lines show the positions and widths of the coronagraphs.}
\end{center}
\end{figure}

\begin{figure}[h!]
\centering
\includegraphics[scale=0.5]{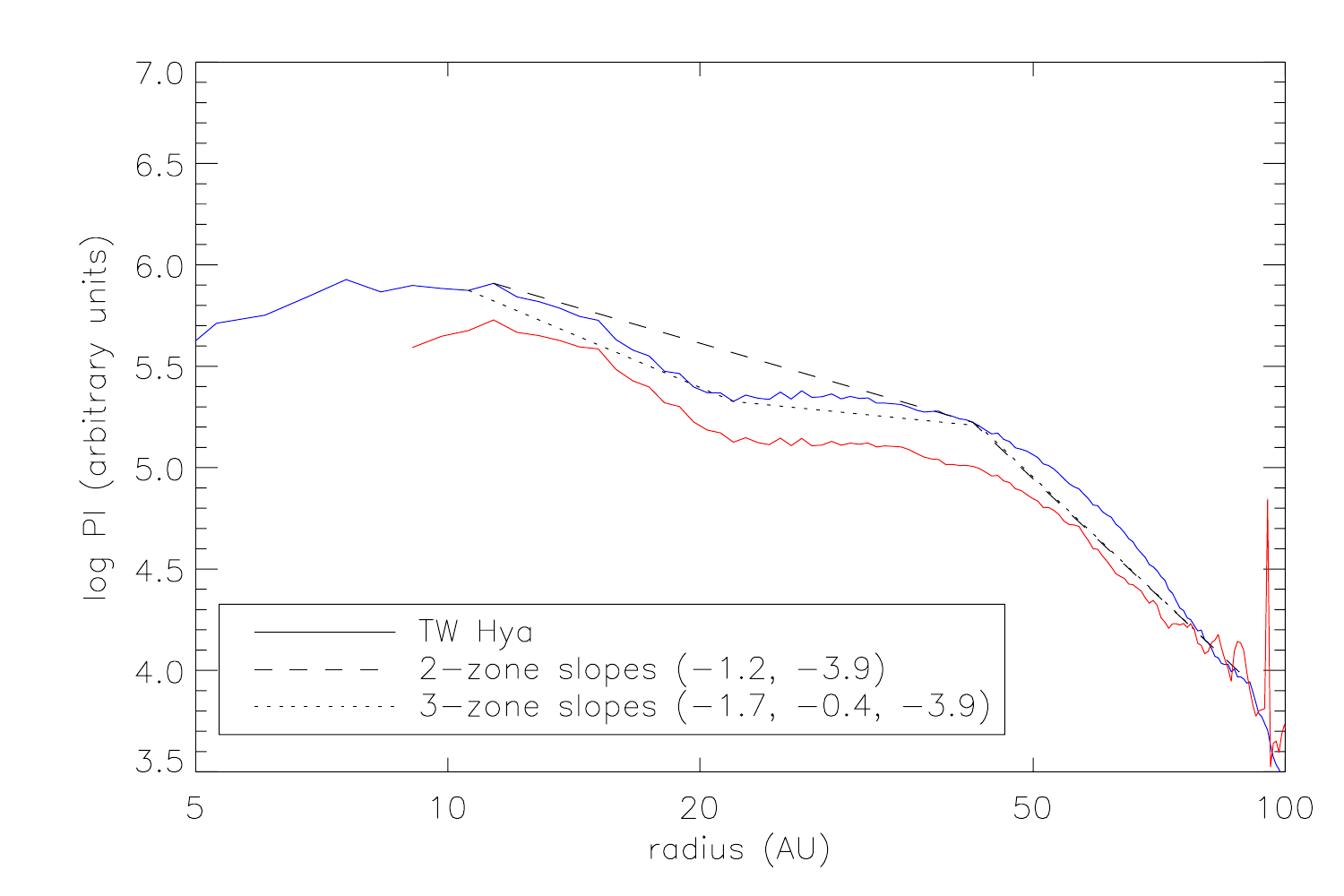}
\includegraphics[scale=0.5]{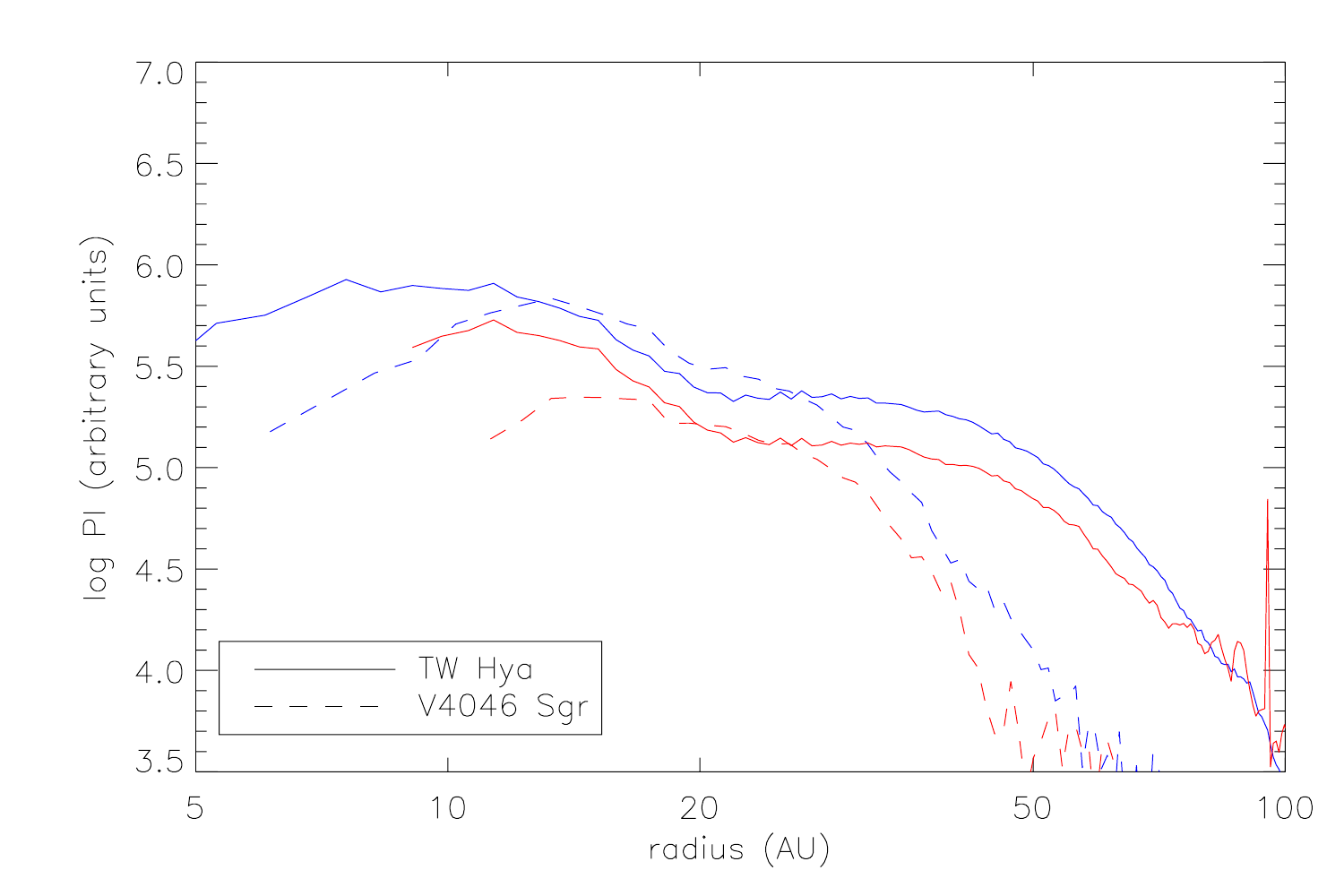}

\caption{\small Left: azimuthally averaged $J$ (blue) and $K1$ (red) radial surface brightness profiles extracted from the GPI images of TW Hya (solid curves), with measured slopes for ``two-radial-zone'' and ``three-radial-zone'' \citep{akiyama2015} models overlaid (black dashed and dotted lines, respectively; see text). Right: comparison of TW Hya radial surface brightness profiles with those of V4046 Sgr \citep[dashed curves; adapted from][]{Rapson2015}. }
\end{figure}

\begin{figure}[h!]
\centering
\includegraphics[scale=0.75]{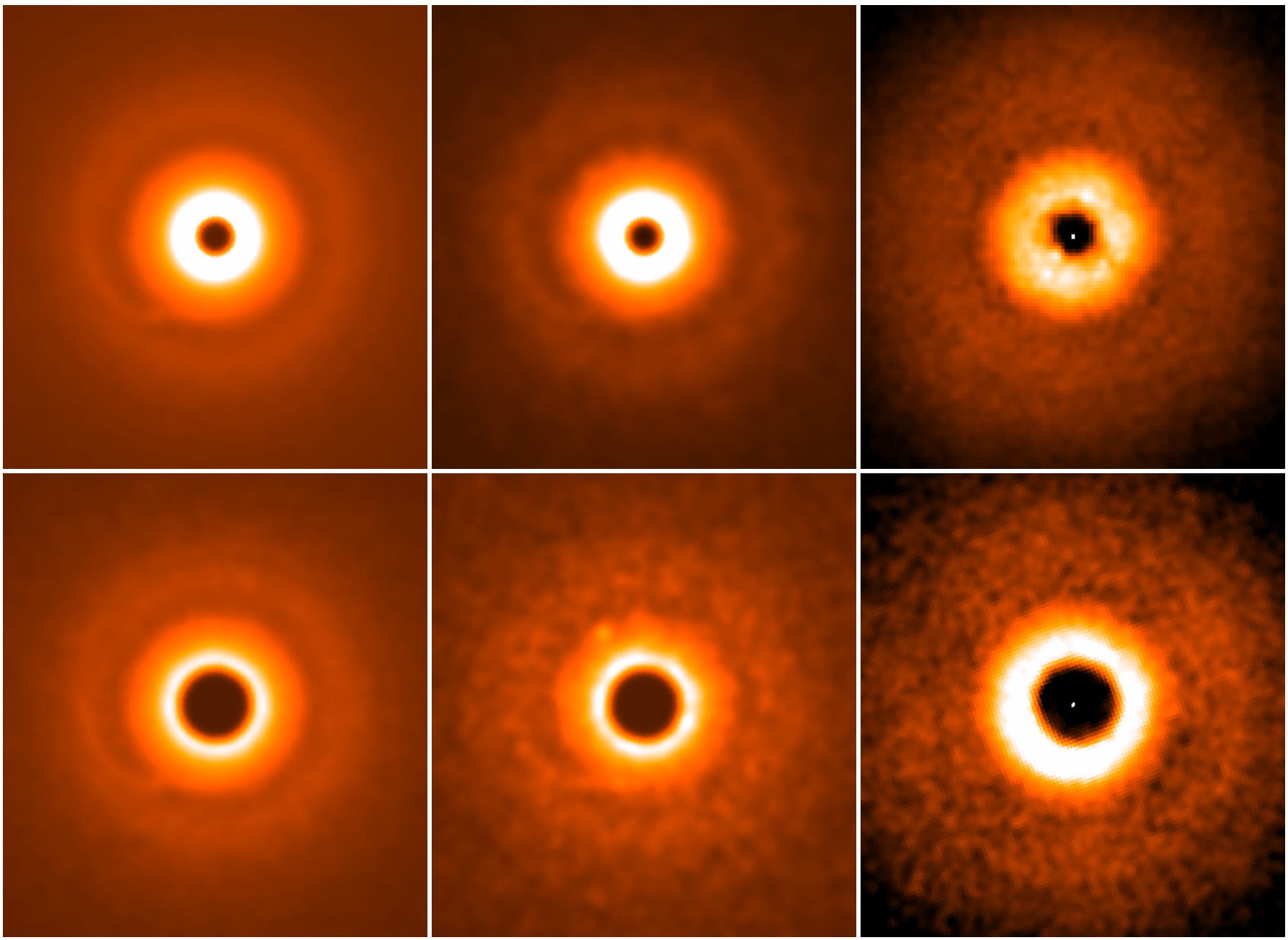}
\caption{\small Comparison of simulated $J$ (top row) and $K1$ (bottom row) polarized intensity (scattered light) images obtained from Monte Carlo models of a disk with embedded 0.16 $M_{Jup}$ planet  \citep[left and middle panels; adapted from][]{Dong2015} with GPI images (right panels). In each row, low-noise and high-noise model images (see text) are presented in the left and middle panels, respectively. All images are displayed on a linear intensity scale and have a field of view of 90 AU $\times$ 90 AU ($1.67''\times1.67''$).}
\end{figure}

\begin{figure}[h!]
\centering
\includegraphics[scale=0.55]{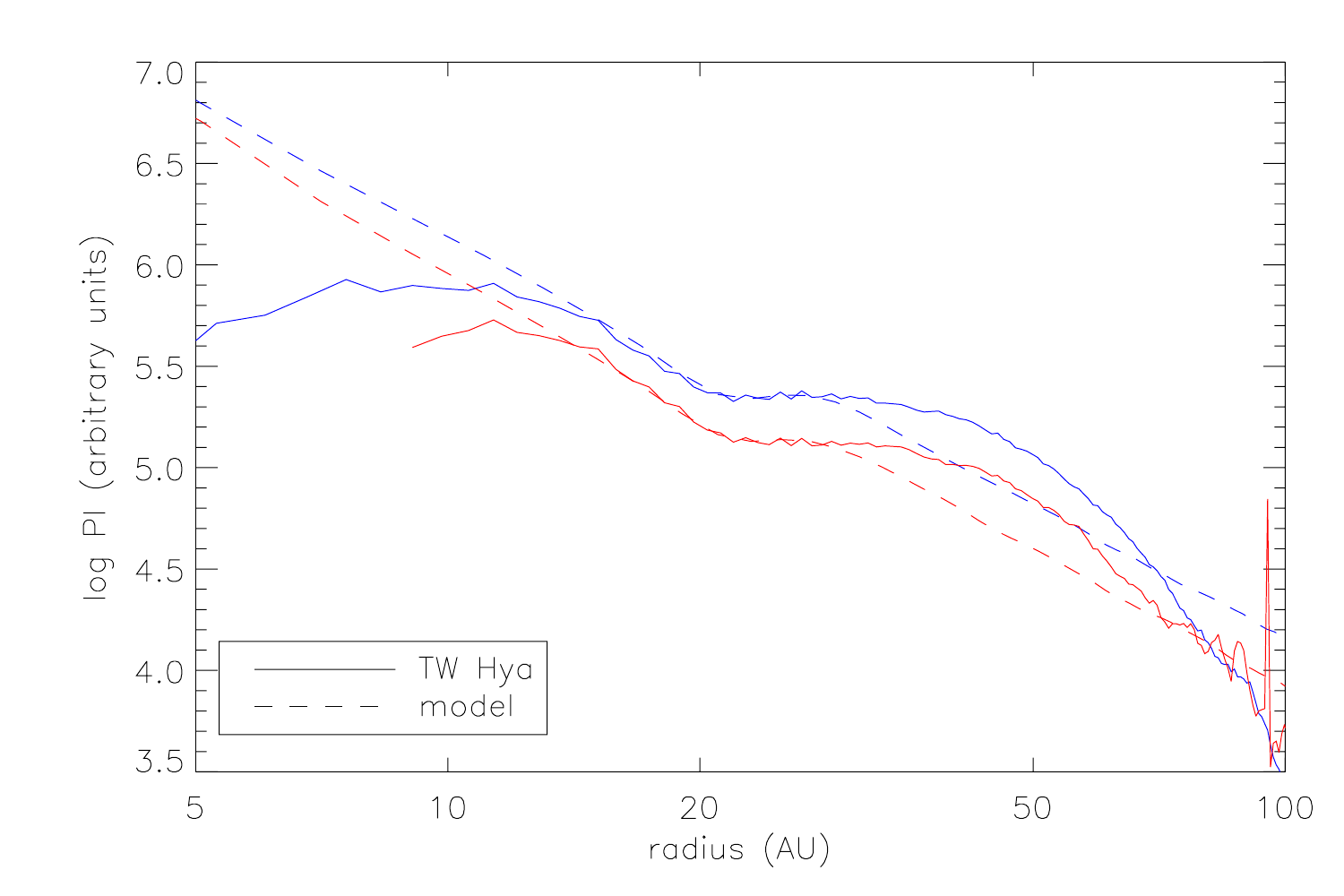}
\caption{\small Radial surface brightness profiles for the TW Hya disk, as in Fig.~3 ($J$: blue; $K1$: red), 
compared with corresponding radial surface brightness profiles obtained from the high-noise model images illustrated in the middle panels of Fig.\ 4.}
\end{figure}

\end{document}